\documentclass[onecolumn,showpacs,amsmath,amssymb,10pt,aps]{revtex4}
\usepackage{amsmath}
\usepackage{amsfonts}
\usepackage{amssymb}
\begin{document}

\title{Bulk Viscosity Effects on the Early Universe Stability}

\author{Nakia Carlevaro and Giovanni Montani}
    \affiliation{ICRA--International Center for Relativistic Astrophysics c/o Dipartimento di Fisica (G9), Universit\`a       di Roma ``La Sapienza'', Piazza A.Moro 5, 00185 Rome, Italy \\and \\Dipartimento di Fisica Universit\`a di Roma          ``La Sapienza'' \\nakia.carlevaro@icra.it\\montani@icra.it}

\date{\today}

\begin{abstract}
We present a discussion of the effects induced by the bulk viscosity on the very early Universe stability. The matter filling the cosmological (isotropic and homogeneous) background is described by a viscous fluid having an ultrarelativistic equation of state and whose viscosity coefficient is related to the energy density via a power-law of the form $\zeta=\zeta_0 \rho^\nu$. The analytic expression of the density contrast (obtained for $\nu=1/2$) shows that, for small values of the constant $\zeta_0$, its behavior is not significantly different from the non-viscous one derived by E.M. Lifshitz. But as soon as $\zeta_0$ overcomes a critical value, the growth of the density contrast is suppressed forward in time by the viscosity and the stability of the Universe is favored in the expanding picture. On the other hand, in such a regime, the asymptotic approach to the initial singularity (taken at $t=0$) is deeply modified by the apparency of significant viscosity in the primordial thermal bath i.e. the isotropic and homogeneous Universe admits an unstable collapsing picture. In our model this feature regards also scalar perturbations while in the non-viscous case it appears only for tensor modes. 
\end{abstract}
\maketitle

\section{Introduction}
The very early stages of the (homogeneous and isotropic) Universe evolution, say at temperatures greater than $\mathcal{O}(10^{16}\,GeV)$, are characterized by a thermal history which can not be regarded as settled down into the equilibrium. Indeed at sufficiently high temperatures the cross sections of the microphysical processes, responsible for the thermal equilibrium, decay like $\mathcal{O}(1/T^{2})$ and they are no longer able to restore the equilibrium while the Universe expands. Thus, going backward, we meet stages where the expansion has an increasing rate and induces non-equilibrium phenomena in the primordial bath. The average effect of having a microphysics unable to follow the Universe expansion by equilibrium stages, results into dissipative processes appropriately described by the presence of bulk (or second) viscosity.

As shown in \cite{bk76,bk77,bnk79} the thermodynamical consequences which bulk viscosity induces on the very early Universe history, can profoundly modify its dynamics and configurations in which matter is created near the singularity become possible. Such scenarios arise because the bulk viscosity coefficient $\zeta$ is a function of the Universe energy density $\rho$ and can be modeled in terms of a power-law as $\zeta=\zeta_0 \rho^{\nu}$ where $\zeta_0,\nu=const$ (for a discussion of this constitution equation see \cite{m73} and see also \cite{m95,dhi01}).

Aim of the present work is to investigate the effects that bulk viscosity has on the stability of the isotropic Universe, i.e. the dynamics of cosmological perturbations is analyzed when viscous phenomena affect the zero and first order evolution of the system. We consider a background corresponding to a Robertson-Walker model filled with ultrarelativistic viscous matter, whose coefficient $\zeta$ corresponds to the choice $\nu=1/2$ and then we develop a perturbation theory which generalizes the Lifshitz works \cite{l46,lk63} to the presence of bulk viscosity. Though the analysis is performed for the case of a flat model, nevertheless it holds in general as soon as the perturbations scale remains much smaller than the Universe radius of curvature. In this respect we deal with perturbations such that $\eta q\ll1$, $2\pi/q$ being the size of the coordinate scale and $\eta$ the conformal time variable. Since the dynamics we consider holds near the singularity for $\eta\ll1$, then we make allowance for arbitrarily large values of $q$ and therefore the condition for the general validity $q\gg2\pi\mathcal{|K|}^{1/2}$ ($\mathcal{K}$ being the Robertson-Walker curvature parameter) can be always fulfilled.

As issue of our analysis we find that two different dynamical regimes appear when viscosity is taken into account and the transition from one regime to the other one takes place when the parameter $\zeta_0$ overcomes a given threshold value. However in both these stages of evolution the Universe results to be stable as it expands; the effect of increasing viscosity is that the density contrast begins to decrease with increasing $\eta$ when $\zeta_0$ is over the threshold. It follows that a real new feature arises with respect to the Lifshitz analysis when the collapsing point of view is addressed. In fact, as far as $\zeta_0$ remains below the threshold value, the isotropic Universe approaches the initial \emph{Big-Bang} with vanishing density contrast and its stability is preserved in close analogy to the non-viscous behavior. But if $\zeta_0$ overcomes its critical value, then the density contrast explodes asymptotically ($\eta\rightarrow0$) and the isotropic Universe results unstable approaching the initial singularity. In the Lifshitz non-viscous analysis this same backward in time instability takes place only when tensor perturbations (gravitational waves) are taken into account, since their amplitude increases backward as the inverse of the cosmic scale factor.

The new feature induced by the bulk viscosity consists of having instability simply in correspondence to scalar perturbations induced by fluctuations in the matter filling the Universe. The cosmological interest in such instability of the primordial Universe (toward scalar perturbations) comes out reversing the picture from collapse to expansion and taking into account the time reversibility of the Einstein equations. In fact if the early Universe does not emerge from the Planck era peaked around the Robertson-Walker geometry (indeed a good degree of generality in its structure is predicted either by classical and quantum argumentation \cite{bm04}) then it can not reach (according to our analysis) an homogeneous and isotropic stage of evolution before the viscous effect become sufficiently small.

Since a reliable estimation \cite{k} fixes the appearance of thermal bath into the equilibrium below temperatures $\mathcal{O}(10^{16}GeV)$ and this limit corresponds to the pre-inflationary age, our result supports the idea that an isotropic universe outcomes only after a vacuum phase transition settled down. In Sec. II we present the paradigm underlying perturbations theory to the Robertson-Walker Universe and provide the key expressions to determine the inhomogeneous corrections corresponding to different states of the matter filling the cosmological space. In Sec. III a brief review of the Lifshitz analysis is presented in order to compare our subsequent results with those ones outcoming from a non-viscous model. In Sec. IV we provide the basic equations governing the perturbations dynamics in a viscous isotropic Universe. In particular we generalize the energy-momentum tensor of the cosmological fluid in view of including the viscous effects and then derive the new zero and first order dynamics; the Friedmann equations corresponding to the viscous case are solved for the choice $\nu=1/2$. In Sec. V we solve the perturbations equations for the same value of the parameter $\nu$; the behavior of the 3-metric perturbations and of the density contrast is provided in the asymptotic limit to the \emph{Big-Bang} and the outcoming issues are discussed. In Sec. VI brief concluding remarks follow and possible upgradings for the description of the bulk viscosity cosmological effects are discussed.

\section{Perturbations theory to the Einstein equations}
In order to describe the temporal evolution of the energy density small fluctuation, we develop a perturbations theory on the Einstein equations. We limit our work to the study of space regions having small dimensions compared with the scale factor of the Universe $a$ \cite{ll}. According to this approximation, we can consider a 3-dimensional Euclidean (time dependent) metric as spatial component of the background line element
\begin{equation}
\label{metric}
ds^2=dt^2 - a^2\,(dx^2+dy^2+dz^2)\;.
\end{equation}
In the linear approximation, perturbed Einstein equations are
\begin{equation}
\label{eqeintpert}
\delta R_\mu^\nu - \frac{1}{2}\delta_\mu^\nu \delta R= 8\pi G \delta T_\mu^\nu\;,
\end{equation}
where the term $\delta T_\mu^\nu$ represents the perturbation of the energy-momentum tensor which describes the properties of the matter involved in the cosmological collapse. The perturbations of the Ricci tensor $\delta R_\mu^\nu$ can be derived from the metric perturbations $h_\mu^\nu=-\delta g_\mu^\nu$, since the general expression for the perturbed curvature tensor is
\begin{equation}
\delta R_{\mu\nu\rho}^{\sigma} =\frac{1}{2}\,(h_{\mu;\,\rho;\,\nu}^{\sigma}+h_{\rho;\,\mu;\,\nu}^{\sigma}-h_{\mu\rho;\,\nu}^{;\,\sigma}-h_{\mu;\,\nu;\,\rho}^{\sigma}-h_{\nu;\,\mu;\,\rho}^{\sigma}+h_{\mu\nu;\,\rho}^{;\,\sigma})\;.
\end{equation} 
For convenience let us introduce a new temporal variable $\eta$, given by the relation $dt=a\,d\eta$, and use the symbol $(')$ for its derivatives; we moreover impose, without loss of generality, that the synchronous reference system is still preserved under perturbations, so that
\begin{equation}
\label{sinc}
h_{00}=h_{0\alpha}=0\;.
\end{equation}
If we consider the background metric (\ref{metric}) the perturbations of the mist components of the Ricci tensor and of the curvature scalar read:
\begin{subequations}
\label{ricci-pert}
\begin{align}
& \delta R_{0}^{0}= - \frac{1}{2a^2}\,h'' - \frac{a'}{2a^3}\,h'\;\,,\qquad\quad
\delta R_{0}^{\alpha}= \frac{1}{2a^2} \,\left(h^{,\,\alpha\,'} -h_{\beta}^{\alpha,\,\beta\,'}\right)\;\,,\\
& \delta R_{\alpha}^{\beta}= -\frac{1}{2a^2}\left(h^{\gamma,\,\beta}_{\alpha,\,\gamma} +h^{\beta,\,\gamma}_{\gamma,\,\alpha}-h^{\beta,\,\gamma}_{\alpha,\,\gamma} -h^{,\,\beta}_{,\,\alpha}\right)-\frac{1}{2a^2}\,h_{\alpha}^{\beta\,''}-\frac{a'}{a^3}\,h_{\alpha}^{\beta\,'}- \frac{a'}{2a^3} \,h'\,\delta_{\alpha}^{\beta}\;\,,\\
& \delta R \,\,\,= -\frac{1}{a^2}\left(h^{\gamma,\,\alpha}_{\alpha,\,\gamma} -h^{,\,\gamma}_{,\,\gamma}\right) - \frac{1}{a^2}\,h'' - \frac{3 a'}{a^3}\,h'\;.
\end{align}
\end{subequations}

By using these expressions we are able to rewrite the left-hand side of Einstein equations through the metric perturbations $h^{\alpha}_{\beta}$ in order to develop the perturbations theory after describing the matter properties via an appropriate energy-momentum tensor.

\section{Primordial Universe in absence of viscosity}
In this Section we assume that the Universe, in its primordial expansion, behaves like a perfect fluid. This hypothesis was made by Lifshitz in his works \cite{l46,lk63} and can be expressed writing the energy-momentum tensor in the form:
\begin{equation}
\label{T-perf}
{T_{\mu}}^{\nu}=\,(\rho+p)u_{\mu}u^{\nu}\,-\,p\,{g_{\,\mu}}^{\nu}\;,
\end{equation}
where $p$, $\rho$ are respectively the pressure a the energy density of the fluid, and $u_\mu$ is its 4-velocity expressed in the comoving system we consider, i.e. 
\begin{equation}
\label{u-comoving}
u^0=1/a \qquad u^\alpha=0\;.	
\end{equation}

Using the synchronous character of the perturbed metric, we are now able to write the perturbations of the above energy-momentum tensor:
\begin{equation}
\label{T-pert}
\delta T_{0}^{0}=\delta \rho\;,\qquad
\delta T_{0}^{\alpha}=a\,(p+\rho)\,\delta u^{\alpha}\;,\qquad
\delta T_{\alpha}^{\beta}=-\delta_{\alpha}^{\beta}\,v_s^2 \delta \rho\;,
\end{equation}
$v_s=\sqrt{\delta p/\delta\rho}\,$ being the sound speed of the fluid. 

Since we use an Euclidean background metric, we can expand the perturbations in plane waves of the form $e^{i\textbf{q}\cdot\textbf{r}}$, where $\textbf{q}$ is the adimensional comoving wave vector being the physical one $\textbf{k}=\textbf{q}/a$. Here we investigate the gravitational stability which is described by the behavior of the energy density perturbation expressible only by a scalar function; thus we have to choose the scalar representation of the metric perturbations in order to involve a change, not only in the gravitational field, but also in the velocity and in the energy density \cite{ll,lk63}. Such a development is made by the scalar harmonics $Q=e^{i\textbf{q}\cdot\textbf{r}}$, from which the following tensor
\begin{equation}
\label{scalar}
Q^{\beta}_{\alpha}=\frac{1}{3}\delta^\beta_\alpha Q\;\,,\qquad
P^{\beta}_{\alpha}=\left[\frac{1}{3}\delta^\beta_\alpha-\frac{q_\alpha q^\beta}{q^2}\right]Q\;\,.
\end{equation}
can be constructed. We can now express the time dependence of the gravitational perturbations through two functions $\lambda(\eta)$, $\mu(\eta)$ and write the tensor $h_\alpha^\beta$ in the form
\begin{equation}
\label{h-form}
h_\alpha^\beta=\lambda(\eta)P_\alpha^\beta+\mu(\eta)Q_\alpha^\beta\;\,,\qquad h=\mu(\eta)Q\;.
\end{equation}

Let us now consider the primordial stages of the Universe expansion, i.e. $\eta\ll1$, when the radiation-like density dominates the matter one. The equation of state is $p=\rho/3$, from which the relations (for a flat Universe ${\cal{K}}=0$) arise
\begin{equation}
\label{r-d-e}
\rho=C a^{-4}\;\,,\qquad a = a_1 \eta\;,\qquad v_s^{2}=1/3\;,	
\end{equation}
where $C$ is an integration constant and $a_1=\sqrt{8\pi GC/3}$. In this approximation we can obtain the basic equations which describe the temporal evolution of the perturbations. Expressing equations (\ref{ricci-pert}) through the representation (\ref{h-form}) and using expressions (\ref{T-pert}) in the form $\delta T_\alpha^\beta=-\delta_\alpha^\beta v_s^2 \delta T_0^0$, the perturbed Einstein equations give, respectively for $\alpha\neq\beta$ and for contraction over these indexes, two equations for the metric perturbations
\begin{equation}
\label{eq-la-mu}
\lambda'' + \frac{2}{\eta}\,\lambda'-\frac{q^2}{3}(\lambda+\mu)=0	\;,\qquad\quad
\mu'' + \frac{3}{\eta}\,\mu'+\frac{2q^2}{3}(\lambda+\mu)=0\;.
\end{equation}
Furthermore, taking the 0-0 components of (\ref{eqeintpert}), we can express the energy density directly from the adopted functions $\lambda$ and $\mu$ in the form
\begin{equation}
\label{rho-form}
\delta\rho=\frac{Q}{24\pi G a^2}\left[q^2(\lambda+\mu)+\frac{3a'}{a}\,\mu'\right]\;.
\end{equation}

Among the solutions there are some which can be removed by a simple transformation of the reference system (compatible with its synchronous character), and therefore they do not represent any real physical change in the metric. The corresponding expression for the metric perturbations can be established, \emph{a priori}, through a coordinates transformation \cite{ll} taking into account the constraint (\ref{sinc}):
\begin{equation}
\label{gauge}
\tilde{h}^{\beta}_{\alpha}=f^{,\,\beta}_{0,\,\alpha}\int\frac{d\eta}{a}\,+\,\frac{a'}{a^2}\,f_0 \,\delta^{\beta}_{\alpha}\,+ \,\left(f^{,\,\beta}_{\alpha}+f^{\beta}_{,\,\alpha}\right)\;\,,
\end{equation}
where $f_0$, $f_\alpha$ are arbitrary (small) functions of the coordinates.

In the assumption $\eta\, q \ll1$ the equations (\ref{eq-la-mu}) admit, in the leading order, the solutions
\begin{equation}
\label{sol-l-mu}
\lambda=\frac{3C_1}{\eta}\,+\,C_2\;,\qquad\;\mu=-\frac{2q^2}{3}\,C_1\,\eta+\,C_2\;,
\end{equation}
where the fictitious solutions (\ref{gauge}), which in our ultrarelativistic approach assume the form $\lambda-\mu=const$ ($f_0=0$, $f_\alpha=P_\alpha$) and $\lambda+\mu\sim1/\eta^2$ ($f_0=Q$, $f_\alpha=0$), are excluded. The final expressions for the gravitational perturbations and for the density contrast $\delta\rho/\rho$ can be obtained substituting this solutions in (\ref{h-form}) and (\ref{rho-form})
\begin{equation}
\label{sol-h}
h^{\beta}_{\alpha}=\frac{3C_1}{\eta}\,P^{\beta}_{\alpha}\,+\,C_2(Q^{\beta}_{\alpha}+P^{\beta}_{\alpha})
\end{equation}
\begin{equation}
\label{sol-rho}
\frac{\delta\rho}{\rho}=\frac{q^2}{9}(C_1\eta+C_2\eta^2)Q	\;.
\end{equation}
Here the constants $C_1$, $C_2$ must satisfy the conditions expressing the smallness of the perturbations at the moment $\eta_0$ when they arise; assuming that harmonics $Q$ are of the unity order magnitude, the inequalities $\lambda\ll1$, $\mu\ll1$ give the constraints $C_1\ll\eta_0\ll1$ and $C_2\ll1$.

The expression of the cosmological perturbation (\ref{sol-rho}) contains terms which increase, in an expanding Universe, proportionally to positive powers of the scale-factor $a=a_1\eta$. This expansion can't, nevertheless, imply the gravitational instability: if we consider the magnitude order $\eta\sim1/q$, the conditions satisfied by the constants $C_1$, $C_2$ imply that the density perturbation remains small even in the higher order of approximation. This behavior of the cosmological fluctuation yields the gravitational stability of the primordial Universe; the only stability we can found in a non-viscous Universe \cite{lk63} is provided by the tensor perturbations $h^{\alpha}_{\beta}$and takes place approaching backward the \emph{Big-Bang}.

\section{Dynamical description of the viscous Universe}
In the last Section we assumed that the Universe could be described by the energy-momentum tensor of an ultrarelativistic perfect fluid. The immediate generalization is to consider the presence of dissipative processes within the fluid dynamics as expected at temperatures above $\mathcal{O}(10^{16} GeV)$; this correction is represented by an additional term in the expression of the energy-momentum tensor (\ref{T-perf}) and it can be derived from thermodynamical properties of the fluid (in particular from the law of increasing entropy  \cite{ll-fluid,e40}). Using the conservation law $T^{\nu}_{\mu;\,\nu}=0$ we arrive \cite{clw92,mont01} to express the energy-momentum tensor of an ultrarelativistic viscous fluid in the form
\begin{equation}
\label{T-visc}
T_{\mu\nu}=(\tilde{p}+\rho)u_{\mu}u_{\nu}-\tilde{p}\,g_{\mu\nu}\;,\qquad\;
\tilde{p}= p - \zeta\,u^{\rho}_{;\,\rho}\;,
\end{equation}
where $p$ denotes the usual thermostatic pressure and $\zeta$ is the \emph{bulk viscosity} coefficient. In this work we neglect the so called shear viscosity (first viscosity) since in the case of isotropic cosmological evolution there is no displacement of the matter layers with respect to each other and this kind of viscosity represents the energy dissipation due to this effect. 

The coefficient $\zeta$ is not constant and we have to express its dependence on the state parameters of the fluid. In the homogeneous models this quantity depends only on time, and therefore we may consider it as a function of the Universe energy density $\rho$. According to literature developments \cite{bk76,bk77,bnk79,m95} we assume that $\zeta$ depends on $\rho$ via a power-law of the form
\begin{equation}
\label{bulk}
\zeta=\zeta_0\,\rho^\nu\;,	
\end{equation}
where $\zeta_0$ is a constant and $\nu$ is an adimensional parameter. The behavior of this parameter is derived by V.A. Belinskii et al. \cite{bk76} for asymptotic values of the density energy yielding the constraint $0\leqslant \nu\leqslant 1/2$ in the region of large $\rho$.

Let us now perturb the viscous energy-momentum tensor in our synchronous reference system obtaining the expressions
\begin{equation}
\label{T-pert-visc}
\delta T_{0}^{0}=\delta \rho\;,\quad
\delta T_{0}^{\alpha}=a\,(\tilde{p}+\rho)\,\delta u^{\alpha}\;,\quad
\delta T_{\alpha}^{\beta}=\delta_{\alpha}^{\beta}\left[-c_s^2 \delta \rho + \zeta\left( \delta u^\gamma_{\,,\gamma}+\frac{h'}{2a^2}\right)\right]\;,
\end{equation}
and the relation between the mist components of the tensor
\begin{equation}
\label{T-pert-1-visc}
\delta T_{\alpha}^{\beta}=\delta_{\alpha}^{\beta}\left[-c_s^2 \delta T_{0}^{0} + \zeta\frac{h'}{2a}+ \frac{\zeta\; \delta T_{0\;\,,\gamma}^{\gamma}}{a(\tilde{p}+\rho)}\right];	
\end{equation}
using the background metric (\ref{metric}) and the expressions (\ref{u-comoving}) of the 4-velocity $u_\mu$ in a comoving system, the viscous pressure $\tilde{p}$ becomes
\begin{equation}
\label{p-prime}	
\tilde{p}= p - 3\,\zeta\,\frac{a'}{a^{2}}\;,
\end{equation}
and the quantity $c_s$ is given by
\begin{equation}
\label{c-sound}
c_s^2\equiv v_s^2 -3 \zeta_0\, \frac{a'}{a^2}\,\nu\rho^{\,\nu-1}\;.
\end{equation} 

Of course the presence of viscosity does not influence the expression of the Ricci tensor and its perturbations, thus we can still keep the expressions (\ref{ricci-pert}) and use the formula (\ref{T-pert-1-visc}) to build up the equations which describe the dynamics of $h_\alpha^\beta$ and $\delta\rho$. It is convenient to choose, as final equations, the one obtained from the Einstein ones for $\alpha\neq\beta$ and for contraction over $\alpha$ and $\beta$, which read respectively  
\begin{equation}
\label{h-1-visc}
\big(h^{\gamma,\,\beta}_{\alpha,\,\gamma} +h^{\beta,\,\gamma}_{\gamma,\,\alpha} -h^{\beta,\,\gamma}_{\alpha,\,\gamma}-h^{,\,\beta}_{,\,\alpha}\big)+ h^{\beta\,''}_{\alpha}\,+\,\frac{{2a'}}{a}\,h^{\beta\,'}_{\alpha}=0\,,\quad\,\alpha\neq\beta\;,
\end{equation}
\begin{equation}
\label{h-2-visc}
\begin{aligned}
\frac{1}{2}\,\big(h^{\gamma,\,\alpha}_{\alpha,\,\gamma}-h^{,\,\gamma}_{,\,\gamma}\big)&\big(1+3c_s^2\big)+h''+\\
&+\frac{a'}{a}\big(2+3c_s^2-12\pi G\frac{a}{a'}\,\zeta\big)\,h'\,+\\ &\qquad\qquad\quad-\frac{3\zeta}{2a(\tilde{p}+\rho)}\big(\,h^{,\,\alpha\,'}_{,\,\alpha}-\,h^{\gamma,\,\alpha\,'}_{\alpha,\,\gamma}\big)=0\;.
\end{aligned} 
\end{equation}

Furthermore the fictitious solutions (\ref{gauge}) stand also in presence of dissipative processes because they are founded by a transformation of synchronous reference system.

Since we want to describe the gravitational instability in presence of viscosity, we consider the scalar representation (\ref{h-form}) which, once substituted in the last expressions, yields the equations describing the perturbations temporal evolution. Let us now express the time dependence of the model variables; as in the last Section, we consider the earlier stages of a flat Universe corresponding to $\eta\ll1$ and with the equation of state $p=\rho/3$. The Universe zero-order dynamics is described by the equation of energy conservation and the Friedmann one, which are respectively
\begin{equation}
\label{continuity-friedmann}
\rho'+3\,\frac{a'}{a}\,(\,\rho+\tilde{p}\,)=0\,,\qquad
\frac{a'}{a^{2}}\,=\,\sqrt{\frac{8}{3}\,\pi G\rho}\;.
\end{equation}
In order to integrate these equations we assume, according to the large energy density of the primordial expansion, $\nu=1/2$  \cite{dhi01,bhm02}. Substituting (\ref{p-prime}) into the above equations we obtain, for $\nu=1/2$: 
\begin{equation}
\label{omega}
\rho=C a^{-(2+2\omega)},\qquad a=a_1\,\eta^{1/\omega},\qquad \omega=1-\sqrt{54\pi G}\;\zeta_0\,,
\end{equation}
being $C$ an integration constant and $a_1=(8\omega^2\pi C G/3)^{1/2\omega}$. Since we consider an expanding Universe, the factor $a$ must increase with positive power of the temporal variable thus we obtain the constraint $0<\omega\leqslant1$.

Using these explicit dependences we get two equations for the $\lambda$, $\mu$ time functions
\begin{equation}
\label{eq-la-visc}
\lambda''+\frac{2}{\omega\eta}\,\lambda'-\frac{q^2}{3}\,\left(\lambda+\mu\right)=0\;,
\end{equation}
\begin{equation}
\label{eq-mu-visc}
\begin{aligned}
\mu''+\Big(\frac{2+3c_s^2}{\omega\eta}\Big)\mu'-\Big(\frac{12 \pi \sqrt{C}G\zeta_0} {a_1^{1+\omega}\,\eta^{1+1/\omega}}&\Big)\mu'+\frac{q^2}{3}\left(\lambda+\mu\right)\left(1+3c_s^2\right)+\\ 
&+\frac{q^2\zeta_0}{4\sqrt{C}/3a_1^\omega-3\zeta_0/\omega}\,\eta\,\left(\mu'+\lambda'\right)=0\,,
\end{aligned} 
\end{equation}
whose solutions describe the evolution of the gravitational perturbations and the energy density fluctuation.

\section{The behavior of density perturbation}
In the model developed in this paper, we study the gravitational collapse dynamics of the primordial Universe near the initial \emph{Big-Bang} for $\eta\ll1$. As in Lifshitz's works, we now analyze the case of perturbations scale sufficiently large to use the approximation $\eta q\ll1$. In the non-viscous model the cosmological stability of the isotropic Universe is guarantied by the positive power-law exponents of the density contrast evolution and by the constraint for the constants $C_1$, $C_2$. In our scheme equations (\ref{eq-la-visc}) and (\ref{eq-mu-visc}) admit asymptotic analytic solutions for the functions $\lambda$ and $\mu$; in the leading order $\lambda$ takes the form
\begin{equation}
\label{sol-l-visc}
\lambda=\frac{C_1}{\eta^{2/\omega-1}}\,+\,C_2\;,
\end{equation}
where $C_1$, $C_2$ are two integration constants. Substituting this expression in (\ref{eq-mu-visc}) we get, in the same order of approximation, the behavior of the function $\mu$ as
\begin{equation}
\label{sol-mu-visc}
\mu=\frac{\tilde{C}_1}{\eta^{1/\omega-3}}\,+\,C_2\;,
\end{equation}
where we have excluded the non-physical solutions (\ref{gauge}) as written in the form $\lambda-\mu=const$. The constant $\tilde{C}_1$ is given by the expression $\tilde{C}_1=A/B(3-1/\omega)$, $A$ and $B$ being constants involved in the equation for $\mu$ having the form
\begin{equation}
\label{A-B}
A=\frac{C_1\,q^2}{3}\left(1+3c_s^2\right)+\frac{C_1(1-2/\omega)q^2\zeta_0}{4\sqrt{C}/3a_1^\omega-3\zeta_0/\omega}\;, \qquad B=\frac{12 \pi \sqrt{C} G\zeta_0}{a_1^{1+\omega}}\;.
\end{equation}

Let us now write the final form of the perturbations pointing out their temporal dependence in the viscous Universe. The gravitational perturbations (\ref{h-form}) become
\begin{equation}
\label{h-visc}
 h^{\beta}_{\alpha}=\frac{C_1}{\eta^{2/\omega-1}}\,P^{\beta}_{\alpha}+
 \frac{\tilde{C}_1}{\eta^{1/\omega-3}}\,Q^{\beta}_{\alpha}+C_2\left(Q^{\beta}_{\alpha}+P^{\beta}_{\alpha}\right)\;,
\end{equation}
and the density contrast reads
\begin{equation}
\label{contr-visc}
\frac{\delta\rho}{\rho}=F_\omega\left[C_1\eta^{3-2/\omega}+C_2\eta^2+C_3\eta^{3-1/\omega}+\tilde{C}_1\eta^{5-1/\omega}\right],
\end{equation}
where $C_3=3A/q^2\omega B$, and $F_\omega=\omega^2Qq^{2}/9$. As in the non-viscous case, we now impose the conditions expressing the smallness of perturbations at the initial time $\eta_0$. The inequalities $h^{\beta}_{\alpha}\ll1$ and $\delta\rho/\rho\ll1$  yield only two fundamental constraints for the integration constants: $C_1\ll\eta_0^{2/\omega-1}$ and $C_2\ll1$ for any $\omega$-value within the interval $(0,\,1]$. Furthermore we find an additional condition which involves the wave number $q$ and the integration constant $C$; in particular a rough estimate for $\omega<1/3$ of the inequalities $\tilde{C}_1\ll\eta_0^{1/\omega-3}$ and $C_3\ll\eta_0^{1/\omega-3}$ yields the condition $q\ll(GC\eta_0)^{-1/2\omega}$ which ensures the smallness of the cosmological perturbations.
 
Using the hypothesis $\eta\ll1$ we can get the asymptotic form the corrections to the cosmological background. The exponents of the variable $\eta$ can be positive or negative according to the value of the viscous parameter $\omega$, which is always positive but less than unity. This behavior produces two different regimes of the density contrast evolution: in the first case $2/3<\omega\leqslant1$, the perturbation increases forward in time which corresponds qualitatively to the same picture of the non-viscous Universe where the density fluctuation increases but remains small; while in the other case $0<\omega<2/3$ the density contrast is suppressed behaving like a negative power of $\eta$. When the density contrast results to be increasing, the presence of viscosity induces a \emph{damping} of the perturbation evolution in the direction of the expanding Universe, so the cosmological stability is fortified since the leading $\eta$ powers are smaller than the non-viscous ones (\ref{sol-rho}).

For $0<\omega<2/3$, i.e. for a large coefficient $\zeta_0$, the density contrast evolution is deeply modify by the presence of viscosity. We get the leading order expression
\begin{equation}
\label{instability}
\frac{\delta\rho}{\rho}=\frac{C_1\,F_\omega}{\eta^{\kappa}}\;,\qquad 
\kappa=2/\omega-3>0\;.	
\end{equation}
This behavior is very different from the non-viscous one since it yields a strong \emph{damping} of the density contrast. In this regime the density fluctuation decreases forward in time but the most interesting result is the instability which the isotropic and homogeneous Universe acquires in the direction of the collapse toward the \emph{Big-Bang}. For $\omega<2/3$ the density contrast diverges when approaching the cosmological singularity, i.e. for $\eta\rightarrow0$. In a non-viscous Universe the only perturbations which are able to generate the same kind of asymptotic instability are the tensor fluctuations (gravitational waves) whereas now the scalar perturbations destroy asymptotically the primordial Universe symmetry. The dynamical implication of this issue is that an isotropic and homogeneous stage of the Universe can not be generated, from generic initial conditions, as far as the viscosity becomes smaller than the critical value associated to the condition $\omega=2/3$.

\section{Concluding remarks}
The main issue of our investigation is to have shown that the isotropic Universe acquires, backward in time, a regime of instability corresponding to sufficiently high values of the viscous parameter $\zeta_0$. Such a window of instability implies that, if the Universe was born sufficiently far from the homogeneous and isotropic stage, than the bulk viscosity (i.e. the absence of a stable thermal equilibrium) works against isotropization mechanisms and the inflation becomes the scenario from which a Robertson-Walker geometry arises (at least on a given scale). The explanation of this result is in the real physical meaning of the bulk viscosity: such viscous effects come out from the difficulty that microphysics finds to restore the thermal equilibrium against the rapid Universe expansion. As a natural consequence of this physical context, bulk viscosity makes unfavored the establishment of an homogeneous stage from a general cosmological dynamics. On the other hand in a Robertson-Walker Universe, already settled down, we expect that, as we find, the viscous effects depress the density contrast because the particles inside the inhomogeneous fluctuations undergo dissipative processes which frozen the growth of the structures. Despite of the reliable feature of our result, the present investigation, as well as the whole previous literature on this subject, relies on a phenomenological ground; in fact the description of the viscous effect is based on the constitutive equation relating the viscosity coefficient to a power-law of the system energy density. This statement appears well-grounded, but nevertheless it requires to be carefully considered in a precise derivation of the viscosity coefficient from a real kinetic theory of matter \cite{ll-kinetic}. We will address for such a point in a further investigation, which will be aimed to yield an upgrading of the present cosmological issue.

\end{document}